\documentclass[aps,prd,preprint,nofootinbib,showkeys,%
showpacs,groupedaddress,byrevtex]{revtex4-1}

\usepackage{bm}
\usepackage{amsmath}
\begin{document}
\title{Macroscopic approach to the Casimir friction force}
\author{V.V.~Nesterenko}
\email[E-mail:~]{nestr@theor.jinr.ru} \affiliation{Bogoliubov Laboratory
of Theoretical Physics, Joint Institute for Nuclear Research,
Dubna 141980, Russia}
\author{A.V.~Nesterenko}
\affiliation{Bogoliubov
Laboratory of Theoretical Physics, Joint Institute for Nuclear
Research, Dubna 141980, Russia}

\date{\today}
\begin{abstract}
The general formula is derived for the vacuum friction
force between two parallel perfectly flat planes
bounding two material media separated by a vacuum gap
and moving relative to each other with a constant
velocity $\mathbf{v}$. The material media are
described in the framework of macroscopic
electrodynamics whereas the nonzero temperature and
dissipation are taken into account by making use of
the Kubo formulae from non-equilibrium statistical
thermodynamics. The formula obtained provides a
rigorous basis for calculation of the vacuum friction
force within the quantum field theory
methods in the condensed matter physics. The revealed
$v$-dependence of the vacuum friction force proves to
be the following: for zero temperature ($T=0$) it is
proportional to $(v/c)^3$ and for $T>0$ this force is
linear in $(v/c)$.
\end{abstract}



\maketitle

\section{Introduction}
The existence of the vacuum friction force is widely discussed over past
time. This force is considered to arise between two perfectly flat planes
bounding material media separated by a vacuum gap and moving parallel to
each other with a constant velocity. A variety of approaches applied to
the study of this subject leads to contradictory results~\cite{pendry, PL,PL-arxiv,%
VP-1,VP}. The issue on hand deals with a stationary, but
irreversible process caused by the dissipation in polarizable media due to
the vacuum friction. Hence, it is worthwhile to study this problem within
the most general approach, i.e., by employing the macroscopic
electrodynamics when describing material medium~\cite{LL} and the Kubo formula for
the linear response of the system to external action~\cite{Kubo}. To the
best of our knowledge such setting of the problem in question has not been
proposed yet.

\section{Underlying formulae}\label{Uf}
We consider a gap of width $l$ between two solid half-spaces, call them 1
$(z<0)$ and 2 $(z>l)$. We assume also that the half-space 1 is at rest in
the laboratory reference frame, and the half-space 2 is moving with a
constant velocity $\mathbf{v}$ which is parallel to the $x$ axis $\mathbf{v}=(v,0,0)$. In what
follows the consideration is conducted only in the laboratory rest frame.

We are interested in the electromagnetic field connected with the
configuration described above. The corresponding Hamiltonian density is
\begin{equation}
\label{e1}
w=\frac{1}{8\pi}\left ( \mathbf{ED} + \mathbf{HB}
\right ).
\end{equation}
The Gaussian units and the notations generally accepted in macroscopic
electrodynamics are used \cite{LL}. The material relations~\cite{LL}
\begin{align}
\mathbf{D}+\frac{\mathbf{v}}{c}\times \mathbf{H}&=\varepsilon\left (
\mathbf{E}+\frac{\mathbf{v}}{c}\times \mathbf{B}
\right ),\nonumber {}\\
\mathbf{B}-\frac{\mathbf{v}}{c}\times \mathbf{E}&=\mu\left (
\mathbf{H}-\frac{\mathbf{v}}{c}\times \mathbf{D}
\right )\label{e2}
\end{align}
enable one to express the displacements $\mathbf{D}$ and $\mathbf{B}$ in
terms of the strength fields $\mathbf{E}$ and $\mathbf{H}$:
\begin{align}
\mathbf{D} &= \varepsilon\mathbf{E} + \kappa\left [1+\varepsilon \mu \left (\frac{\mathbf{v}}{c}
\right )^2\right ]\left (\frac{\mathbf{v}}{c}\times \mathbf{H}
\right )
-\varepsilon\kappa\left[
\frac{\mathbf{v}}{c}\left(\frac{\mathbf{v}}{c}\cdot\mathbf{E}\right) -
\left(\frac{\mathbf{v}}{c}\right)^{\!2}\mathbf{E}\right]\!,\nonumber {}\\
\mathbf{B} &= \mu\mathbf{H} - \kappa \left [1+\varepsilon \mu \left (\frac{\mathbf{v}}{c}
\right )^2\right ]
\left (\frac{\mathbf{v}}{c}\times \mathbf{E}
\right )
-\mu\kappa\left[
\frac{\mathbf{v}}{c}\left(\frac{\mathbf{v}}{c}\cdot\mathbf{H}\right) -
\left(\frac{\mathbf{v}}{c}\right)^{\!2}\mathbf{H}\right]\!,
\label{e2a}
\end{align}
where $\kappa =\varepsilon \mu -1$. We restrict ourselves to the accuracy
up to $(v/c)^3$ inclusively. As a result the Hamiltonian
density~(\ref{e1}) assumes the form:
\begin{equation}
w=w_0-\frac{\kappa}{c^2}\left [1+\varepsilon \mu \left (\frac{\mathbf{v}}{c}
\right )^2\right ]\left ( \mathbf{v\cdot S}
\right )+
\frac{\kappa}{8\pi}\left (\frac{v}{c}\right )^2
\left [ \varepsilon (E_y^2+E_z^2) + \mu (H_y^2+H_z^2)
\right ], \label{e3}
\end{equation}
where $\mathbf{S}=(c/4\pi)(\mathbf{E}\times \mathbf{H})$ is the
Poynting vector and
\begin{equation}
\label{e4}
w_0=\frac{1}{8\pi}\left (\varepsilon  {\mathbf{E}}^2 + \mu{\mathbf{H}}^2
\right ).
\end{equation}
Obviously the terms in (\ref{e3}) depending on the velocity $\mathbf{v}$
of a medium concern  only  the half-space 2; in the case of the half-space
1 and the vacuum gap $0<z<l$ these terms are absent. For simplicity we
consider both media to be identical.

The motion of the half--spaces relative to each other will be treated as a
weak perturbation that allows us to use the linear response
theory~\cite{Kubo}. In this approach, it is supposed that in the remote
past $(t\to -\infty)$ the perturbation was absent $(\mathbf{v}=0)$ and the
whole system (electromagnetic field in both media 1,\, 2 and in the gap)
was in thermodynamical equilibrium at the temperature $T$. Then the weak
perturbation
\begin{align}
\overline{w} = w - w_0
=&-\frac{v}{c}\left [
1+ \varepsilon \mu \left (\frac{v}{c}\right )^2
\right ]\frac{\kappa}{c}S_x
+\frac{\kappa}{8\pi}\left (\frac{v}{c}\right )^2
\left [ \varepsilon (E_y^2+E_z^2) + \mu (H_y^2+H_z^2)
\right ]\nonumber \\
=& \,\frac{v}{c}\left [
1+ \varepsilon \mu \left (\frac{v}{c}\right )^2
\right ]w_1+\left (
\frac{v}{c}
\right)^2w_2
\label{e4a}
\end{align}
 is switched on adiabatically, so
that the system does not go far away from the initial equilibrium state.
When considering this state we are dealing with the electromagnetic field
described by the Hamiltonian density $w_0$ at the temperature $T$. This
field is connected with unbounded medium at rest possessing the vacuum gap
of the width $l$ made up by two parallel planes. Thus the unperturbed
state is a standard Lifshitz configuration at the temperature~$T$.

As the response of the system under study to the
perturbation~$\overline{w}$ we consider the ponderomotive
force acting on the medium~2 along the~$x$ axis. Obviously
it is the vacuum friction force in the problem on hand. The
density of this force
\begin{equation}
\label{e5} \sum _{\beta=x,y,z}\frac{\partial \sigma_{x\,
\beta}(\bm{r})}{\partial r_\beta}
\end{equation}
is given by the Maxwell stress tensor~\cite{LL}
\begin{equation}
4\pi\sigma _{\alpha \beta}= E_\alpha D_\beta+
H_\alpha B_\beta
-\frac{\delta_{\alpha \beta}}{2}\left [(\mathbf{E \cdot D})
+(\mathbf{H\cdot B})
\right ], \quad
 \alpha, \beta =x,y,z\,{.} \label{e6}
\end{equation}
For the sake of simplicity we use the Minkowski nonsymmetric
form of the stress tensor $\sigma_{\alpha \beta}$ (see
ref.\ \cite{LL,Pauli}). This point is not crucial for us
here. We shall turn to the Abraham symmetric
energy-momentum tensor later.

Taking into account the geometry of the configuration under consideration
it is enough to consider only the component
\begin{equation}
\label{e6a}
\sigma_{xz}=\frac{1}{4\pi}(E_x D_z+H_xB_z)
\end{equation}
on the plane bounding the half-space 2, i.e.,
$\sigma_{xz}(\bm{r}_0)$, where $\bm{r}_0=(x,y,z=l-0)$. This quantity is the
tangential strength in the $x$-direction exerted to
the unit area of the surface $z=l-0$.

By making use of the solution to the material relations~(\ref{e2a}) we
obtain in the $(\mathbf{v}/c)^3$-approximation the following formula for
$\sigma_{xz}$:
\begin{align}
4\pi \,\sigma _{xz}&=
\left[1+\kappa\left(\frac{v}{c}\right)^{\!2\,}\right]
(\varepsilon E_x E_z +\mu H_x H_z)
+\kappa \frac{v}{c}\left [
1+\varepsilon \mu \left (
\frac{v}{c}
\right )^2
\right ](E_x H_y-E_yH_x)\nonumber \\
&=\left[1+\kappa\left(\frac{v}{c}\right)^{\!2\,}\right]4\pi\,\sigma^{(0)}_{xz}+\frac{v}{c}
\left[1+\varepsilon \mu\left(\frac{v}{c}\right)^{\!2\,}\right]4\pi\,\sigma_{xz}^{(1)}{.} \label{e9}
\end{align}
As in the Hamiltonian density~(\ref{e3}), the terms depending on
$\mathbf{v}$ in (\ref{e9}) are relative only to the medium~2.

\section{The Kubo formalism}

Now we are in position to write out the general formula for
the vacuum friction force in the framework of the Kubo formalism~\cite{Kubo}
\begin{equation}
\label{e14}
\langle\sigma_{xz}(\bm{r}_0)\rangle={\langle\sigma_{xz}(\bm{r}_0)\rangle}_0
+\int\limits_{-\infty}^{+\infty}dt'
\int d\bm{r}\,\langle\!\langle
\sigma_{x z}(t,\bm{r}_0),\overline{w}(t',\bm{r})
\rangle\!\rangle\,
{.}
\end{equation}
The integration over $d\bm{r}$ is carried out only in the
half-space 2, i.e., for $z\geq l$ (cf.~with ref.~\cite{Brevik}).
Here $\langle\, ... \,\rangle_0 =\text{Tr}(\varrho_0\,...)$ denotes
the averaging with the Gibbs statistical
operator $\varrho_0=\exp[(F-H_0)/kT]$, where
\begin{equation}
\label{e15}
H_0=\int d\bm{r}\, w_0(\bm{r})
\end{equation}
and the integration is carried out over the both media
1, 2 and over the vacuum gap; $F$~is the corresponding
free energy. The brackets
$\langle\, ... \,\rangle=\text{Tr}(\varrho\,...)$ stand for the
analogous  averaging with the statistical operator
$\varrho$ which takes into account, in addition to
$w_0$, the quantum-mechanical perturbation
$\overline{w}$, the latter being treated in the linear
approximation. The density operator $\varrho$ obeys the condition
$\varrho \to \varrho_0$, when $t\to -\infty$. The main object in~(\ref{e14})
is the retarded Green function\footnote{Kubo~\cite{Kubo} used the
response function $\varphi_{AB}(t-t')$ which is related to the Green
function in a simple way
$
\big\langle\!\big\langle A(t),B(t') \big\rangle\!\big\rangle =
-\theta(t-t')\varphi_{AB}(t-t')\,{.}
$}
for two operators $A(t)$ and $B(t')$:
\begin{equation}\label{e16}
\big\langle\!\big\langle A(t),B(t') \big\rangle\!\big\rangle =
\frac{1}{i\hbar}\theta(t-t')\big\langle\!\bigl[A(t),B(t')\bigr]\!\big\rangle_0{.}
\end{equation}
 The explicit time dependence of the operators $A(t)$ and
$B(t')$ in (\ref{e16}) implies the Heisenberg representation with the
Hamiltonian $H_0$ from (\ref{e15}).

It is easy to show that the first term in the right-hand side of
(\ref{e14}) vanishes $\langle\sigma(\bm{r}_0)\rangle_0=0$. Indeed
\begin{align}
4 \pi\,\sigma^{(0)}_{xz}&= \varepsilon E_x E_z+\mu H_x H_z=4 \pi
\left .\sigma _{xz} \right |_{\,\mathbf{v}=0}{,} \label{e17}\\
4 \pi\,\sigma^{(1)}_{xz}&= \kappa(E_x H_y-E_y H_x)\,{.} \label{e18}
\end{align}
At the equilibrium state the expected values of all
components of the Maxwell stress tensor  should
vanish, hence $\big\langle\sigma^{(0)}_{xz}\big\rangle_0=0$. The value of
$\big\langle\sigma^{(1)}_{xz}\big\rangle_0$ also vanishes because
$\mathbf{E}(t,\bm{r})$ and
$\mathbf{H}(t,\bm{r})$ are not correlated at the same time $t$ and point
$\bm{r}$~\cite{LLstat2}.

Now we can simplify the integral term in (\ref{e14}).
We take into account the fact that only the terms of
the odd power in $(v/c)$  can be put down to the
friction force. As a result we obtain in the
$(v/c)^3$-approximation
\begin{align}
\langle\sigma_{xz}(\bm{r}_0)\rangle=&\,\frac{v}{c}
\left [1+(\kappa +\varepsilon \mu)\left ( \frac{v}{c}
\right )^2
\right ]\int \limits_{-\infty}^{+\infty} dt'\int d\bm{r}
\,\big\langle\!\big\langle
\sigma_{xz}^{(0)}(t,\bm{r}_0),w_1(t',\bm{r})
\big\rangle\!\big\rangle
\nonumber {} \\
&+\left (
\frac{v}{c}
\right )^3 \int \limits_{-\infty}^{+\infty} dt'\int d\bm{r}
\,\big\langle\!\big\langle
\sigma_{xz}^{(1)}(t,\bm{r}_0),w_2(t',\bm{r})
\big\rangle\!\big\rangle
{,}
\label{e19}
\end{align}
where $\sigma_{xz}^{(0)},\; \sigma_{xz}^{(1)}$ are defined in (\ref{e9}), (\ref{e17}),
 (\ref{e18}) and $w_1,\;w_2$ are introduced in (\ref{e4a}) and read
\begin{equation}
\label{e20}
w_1=-\frac{\kappa}{c}S_x,\quad w_2=\frac{\kappa}{8\pi}[\varepsilon (E_y^2+E_z^2)+\mu
(H^2_y+H_z^2)]\,{.}
\end{equation}

Let us prove that the first term in (\ref{e19})
vanishes at zero temperature. First of all it is to be
noted that at $T=0$ and $\mathbf{v}=0$ the
electromagnetic field in the half-space 2 can be
considered as  isolated system with its conserved
total momentum  $P_x$:
\begin{equation}
\label{e21} P_x=\frac{1}{c}\int\limits_{z\geq
l}d\bm{r}T^{x0}(\bm{r}) =\frac{1}{c}\int\limits_{z\geq
l}d\bm{r}T^{0x}(\bm{r})=\frac{1}{c^2}\int\limits_{z\geq
l}d\bm{r} S_x(\bm{r})\,{.}
\end{equation}
Here we have assumed that $T^{x0}=T^{0x}$ are the
components of the Abraham symmetric energy-momentum
tensor \cite{Pauli,LL2}. With regards for this the
commutator entering the first term in (\ref{e19})
gives (see, for example \cite{Schwinger}):
\begin{equation}
\label{e22} \int\limits_{z\geq
l}d\bm{r}\big\langle[\sigma^{(0)}_{xz}(t,\bm{r_0}),S_x(t',\bm{r})]\big\rangle_0=
c^2\big\langle[\sigma^{(0)}_{xz}(t,\bm{r_0}),P_x]\big\rangle_0=
c^2\frac{\hbar}{i}\frac{\partial}{\partial x}
\big\langle\sigma^{(0)}_{xz}(t,\bm{r_0})\big\rangle_0=0\,{.}
\end{equation}
In this formula the averaging $\langle\, ... \,\rangle_0$ is carried out
with respect to the lowest energy state, i.e., with
respect to the quantum-field vacuum state.

For $T>0$ and $\mathbf{v}=0$ this reasoning does not
hold because in this case the electromagnetic field in
the half-space 2  interacts with the black body
radiation filling the  gap $0<z<l$. It is this
interaction that ensures the equilibrium of
electromagnetic field in half-spaces 1,2 and in the gap.

Thus for $T>0$ the vacuum friction force is defined by
the first  term in (\ref{e19}) which is linear in
$(v/c)$. For $T=0$ this force is described by the
second term in (\ref{e19}) which is proportional to
$(v/c)^3$. It should be noted that in our approach
this dependence of the Casimir friction force on the
relative velocity has, as a matter of fact, the
kinematical reason.   In refs.\
\cite{pendry97,Barton,Brevik2} such $v$-dependence of the
vacuum friction force was  obtained in a simple quantum mechanical models.

\section{Discussion and conclusion}
Further calculations demand construction of the
four--point Green functions entering final formula
(\ref{e19}), i.e., expressing them in terms of the basic
two--point retarded Green function of electromagnetic
field in a medium~\cite{LLstat2}. This procedure can
wittingly be brought about for linear dielectrics
\cite{Agarwal}. However its realization and removing
the divergencies is a nontrivial task which requires
development of special technique and approximation.
All this is beyond the scope of the present paper.

Closing we would like to stress the following. The
final formula (\ref{e19}) provides a rigorous basis for
calculation of the vacuum friction force in  the
framework of the quantum field theory methods in the
condensed matter physics. The revealed $v$-dependence
of the vacuum friction force proves to be the
following: for zero temperature ($T=0$) it is
proportional to $(v/c)^3$ and for $T>0$ this force is
linear in $(v/c)$.

It is also important to note that the Green functions in
resulting formula (\ref{e19}) involve only unperturbed
electric ($\mathbf{E}$) and magnetic ($\mathbf{H}$)  fields
governed by ``free'' Hamiltonian $H_0$ describing the
standard Lifshitz configuration (see Sec.\ \ref{Uf}). Thus
in our approach, unlike other considerations of this
problem (see, for example, \cite{VP-1} and references therein)
there is no need to solve the Maxwell equations with moving boundaries.

Another obvious advantage of the proposed approach  to the calculation of
the Casimir friction force is a correct treatment, from the very
beginning, of the relativistic invariance in this problem~\cite{Barton}.
It is due to the employment of the Minkowski material relation (\ref{e2}).

\begin{acknowledgments}
The authors are thankful to A.\,A.\ Starobinsky for
reading the paper and for useful comments.
The discussions of the subject under study with G.\
Barton and I.\ Brevik were helpful.
 VVN acknowledges the financial
support of the Russian Foundation for Basic Research
Grant No.\ 11-02-12232-ofi-m-2011.
\end{acknowledgments}

\end{document}